\documentclass[aps,pra,twocolumn, showpacs]{revtex4}
\usepackage{graphicx}
\usepackage{dcolumn}
\usepackage{bm}

\begin{document}
\preprint{preprint}
\title{Collective Atomic Motion in an Optical Lattice formed inside a High
Finesse Cavity}
\author{B. Nagorny}
\author{Th. Els\"{a}sser}
\author{A. Hemmerich}
\affiliation{Institut f\"{u}r Laser--Physik, Universit\"{a}t Hamburg,
Luruper Chaussee 149, D--22761 Hamburg, Germany}

\date{\today}

\begin{abstract}
We report on collective non-linear dynamics in an
optical lattice formed inside a high finesse ring cavity in a so far unexplored
regime, where the light shift per photon times the number of trapped atoms
exceeds the cavity resonance linewidth. We observe bistability and
self--induced squeezing oscillations resulting from the retro--action of
the atoms upon the optical potential wells. We can well understand most of
our observations within a simplified model assuming adiabaticity of the
atomic motion. Non-adiabatic aspects of the atomic motion are reproduced
by solving the complete system of coupled non--linear equations of motion.
\end{abstract}


\maketitle

Single atoms interacting with a few photons inside a high finesse
cavity are an extensively studied key system of quantum optics.  Only
recently the significance of the atomic motion in this model system has
been recognized and a wealth of new physics was found including cooling and
trapping single atoms by single photons \cite{Pin:00, Hoo:00}.  More
recently, it has been pointed out that optical cavities with sufficient
finesse and large mode volumes could provide a new means to cool and trap
even large atomic samples by coherent scattering, with the promise of
extending laser cooling to a wider class of atomic species and possibly
even molecules \cite{Doh:97, Hor:97, Hec:98, Cha:03, Els:03}.  A so far
unexplored regime of atom--cavity interactions arises for large atom
samples trapped in cavity--enhanced far off--resonant light fields, if the
collective coupling strength (i.e. the light shift per photon times the
atom number) exceeds the cavity resonance linewidth.  For sufficiently high
finesse of the cavity, the otherwise tiny retro--action of the moving atoms
upon the light field becomes a significant feature of this system.  This
leads to an inherently collective character of the atomic motion and a
corresponding non--linear dynamics of the intra--cavity field.  Such
collective long--range interactions might allow to create controlled
quantum entanglement \cite{Hem:99}, a perspective possibly useful for
quantum computation with neutral atoms.  Other intriguing perspectives are
the implementation of improved cooling schemes for atomic species otherwise
not accessible, for example, as indicated in ref.  \cite{Els:03}.
Sympathetic cooling without collisions might become possible for low
particle densities via thermalization of atom samples trapped at distant
locations inside the cavity.

In this paper we explore the collective motion of atoms
trapped inside an optical standing wave formed by two mutually counterpropagating travelling
wave modes in a high--finesse ring resonator.
A fast servo tightly locks the laser frequency in resonance with one of the modes.
Within a narrow window around the case of exactly symmetric pumping of both cavity modes, a stable optical
lattice is formed. Asymmetric pumping on the order of a few percent yields
surprisingly complex dynamics.
The most striking feature is a drop of the intra--cavity intensity of the unlocked
mode by more than an order of magnitude, if the incoupled powers deviate by
only a few percent in favour of the locked mode.
This reduced intensity level is maintained up to several ten milliseconds
until the slowly decaying lattice population and thus the collective
interaction strength fall below a certain value, where the system
suddenly jumps back to the mode of operation characteristic for a cavity
without atoms. This jump is accompanied by a sudden spatial shift of the
lattice. Depending on the total intensity directed to the cavity and the exact
degree of pumping asymmetry the bistable behavior can be more or less pronounced
and under certain circumstances self--induced radial squeezing oscillations
of the atoms arise. Non--linear dynamics due to optical pumping and
saturation is known to occur in low--finesse resonators
operating close to an atomic resonance \cite{Lam:95}.
Our lattice operates far from resonance where optical pumping and
saturation are negligible and the atoms merely act as a dispersive medium.

Our system is described in detail in ref. \cite{Nag:03} and sketched in
Fig.1(a). The triangular ring cavity has a finesse of $1.8
\times 10^5$, a cavity resonance linewidth of $\gamma_c / \pi = 17.3~kHz$
and a large mode volume of 2.6~$mm^3$. The linearly polarized output of a diode
laser is servo locked into resonance with one of the travelling wave modes.
The laser frequency is red detuned with respect to the D2-line of rubidium ($^{85}Rb$) by
0.7~$nm$. Cold rubidium atoms are provided by a standard magneto-optic
trap (MOT). We can observe the powers transmitted through the cavity and perform
temperature and position measurements by means of fluorescence imaging.
Our intra--cavity optical lattice operates in the regime of strong collective interactions
characterized by $U N \geq 1$, where $N$ is the number of trapped
atoms and $U \equiv \Delta_0 / \gamma_c$, with
$\Delta_0 = 0.091 s^{-1}$ being the light shift per photon.

\begin{figure}
\includegraphics[scale=0.48]{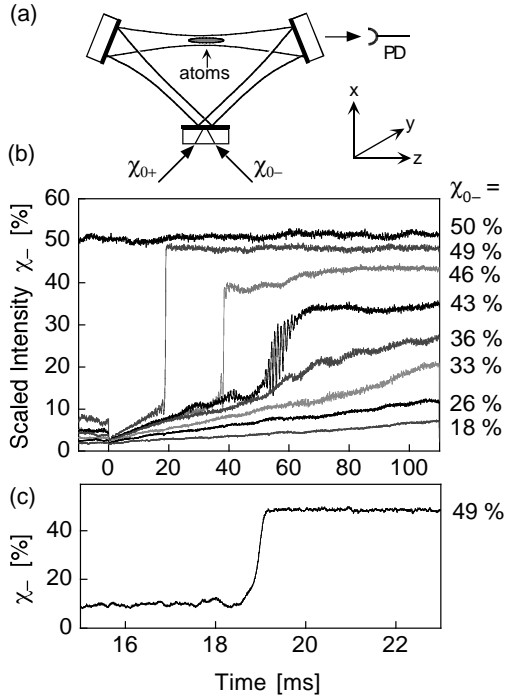}
\caption{ \label{Fig1} (a) Sketch of experimental system, PD = photo diode.
(b) Transmitted intensity of unlocked mode for
$\chi_{0\pm} = 50 \%, 49 \%, 46 \%,43 \%, 36 \%, 33 \%, 26 \%, 18 \%$.
(c) Expanded view of the jump feature in the $49 \%$ trace of (b).}
\end{figure}

In a typical experimental cycle the MOT superimposed on the lattice loads up to several $10^6$
atoms into the lattice, before it is shut off and the atoms are kept by the lattice only.
We record the power of the unlocked mode leaking out of the cavity through one of the mirrors
as illustrated in Fig.1(a). From the known transmission of the mirror we obtain the
intra-cavity intensity $I_{-}$  of the unlocked mode. Similarly we can
observe the intra-cavity intensity $I_{+}$ of the locked mode.
We thus measure the scaled quantities $\chi_{\pm} \equiv I_{\pm}/I_{0}$, where
$I_{0}$ denotes the sum of the intensities in both travelling modes without atoms inside the cavity.
We provide adjustable powers $P_{0\pm}$ to be coupled into the travelling wave modes
which we refer to in terms of the scaled quantities $\chi_{0\pm} \equiv P_{0\pm}/P_{0}$ with
$P_{0}$ being the total power coupled into the cavity. Note that $\chi_{\pm} =
\chi_{0\pm}$, if no atoms are present inside the cavity. Moreover the fast
servo acts to maintain the condition $\chi_{+} = \chi_{0+}$ for any value of
the atom--cavity coupling $U N$.

In Fig.1(b) we illustrate the time evolution of the intra--cavity
intensity in the unlocked mode. In the uppermost trace we have adjusted perfect symmetric
pumping, i.e, $\chi_{0\pm}=50 \%$. In this case we find a constant value
$\chi_{-}=50 \%$ independent of the presence of atoms or the MOT beams which
are shut off at $t=0$~ms. Both modes build up with equal intensities and a stable optical lattice
is formed. This is confirmed by temperature measurements,
showing that the atoms are kept for long times at the initial temperature
provided by the MOT. If $\chi_{0-}$ is adjusted above $50 \%$, $\chi_{-}$
initially drops to $50 \%$ independent of the exact value of $\chi_{0-}$.
As the atom number $N$ decreases in time $\chi_{-}$ gradually approaches $\chi_{0-}$.
For the locked mode $\chi_{+} = \chi_{0+}$ is maintained at all times.
More dramatic consequences arise for values of $\chi_{0-}$ below $50 \%$.
The lower traces in Fig.1(b) are observed
for $\chi_{0-}$ adjusted to $49 \%, 46 \%, 43 \%, 36 \%,
33 \%, 26 \%$, and $18 \%$ respectively. The corresponding values of the initial interaction
strength $U~N(t=0) \approx$ 4.48, 4.25, 4.01, 3.54, 3.30, 2.95, 2.48 are
carefully determined by measuring
the initial particle number $N(t=0)$ via fluorescence detection.
The observed decrease of $N(t=0)$ with decreasing $\chi_{0-}$ arises because the capture
efficiency decreases with the lattice well depth.
Note that already during the MOT phase
a significantly reduced value of $\chi_{-}$ is found. When the MOT is turned
off, $\chi_{-}$ drops to nearly zero. Subsequently, as the number of trapped atoms
decreases due to collisional losses, $\chi_{-}$
gradually increases until a sudden jump occurs which
nearly restores the intensity level $\chi_{0-}$ expected in absence of atoms.
 The time for this jump to occur increases with decreasing $\chi_{0-}$ while the level
to which the intensity initially drops decreases. Moreover, as $\chi_{0-}$
decreases, the jump feature washes out and is not anymore visible below
$\chi_{0-} = 38 \%$. In the vicinity of the boundary, where the jump feature
vanishes, pronounced oscillations in the kHz range are observed,
corresponding to twice the radial vibrational frequency.
An example for this behavior is shown in the $43 \%$ trace of Fig.1(b).
An expanded view on the jump in the
$\chi_{0-} = 49 \%$ trace of Fig.1(b) is given in Fig.1(c) showing some residual
oscillatory behavior right before the jump. Note that the observed rise time
approximately equals a quarter of the radial oscillation time. The observed drop in
transmission for the unlocked mode in the traces of Fig.1(b) is accompanied by a
corresponding increase of the reflected power, i.e.,
the overall intra--cavity intensity is reduced in the presence of atoms.
Note that in all cases the unperturbed cavity condition $\chi_{+} = \chi_{0+}$ is
well kept for the locked mode. We have measured the radial temperature before and after the
jump finding T = 90~$\mu$K and T = 142~$\mu$K respectively. The temperature increase
observed is compatible with adiabatic heating
expected as a result of the change in potential well depth during the
jump. This is plausible since the observed rate of change of the
potential well depth satisfies the condition for adiabaticity.

\begin{figure}
\includegraphics[scale=0.48]{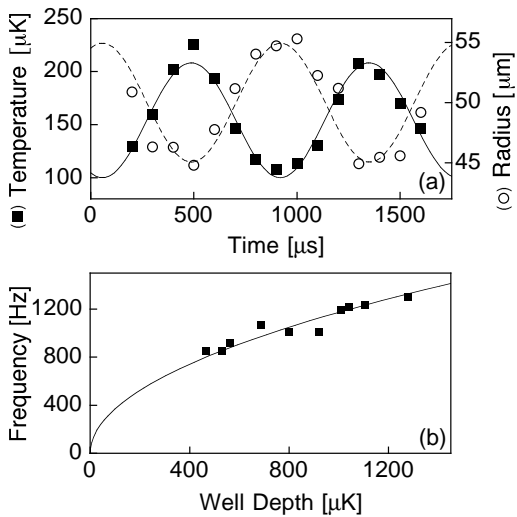}
\caption{ \label{Fig2} (a) Oscillation of the atomic position
(circles) and momentum spread (rectangles). The solid lines are
trigonometric fits with $\pi$ phase delay. (b) Oscillation
frequency plotted versus the well depth. The solid line shows the
expected square root dependence.}
\end{figure}

In order to understand the oscillation feature in Fig.1(b) we have measured
the atomic position and momentum distributions during this oscillation. The
results are plotted in Fig.2(a). We observe an anticyclic variation of the
position and momentum spread at twice the radial vibrational frequency.
This shows that the origin of the oscillatory feature is a self--induced squeezing
oscillation arising for the radial degrees of freedom.
This explanation is confirmed by plotting the observed
oscillation frequency versus the well depth as shown in Fig.2(b) where the expected
square root dependence is found.

\begin{figure}
\includegraphics[scale=0.48]{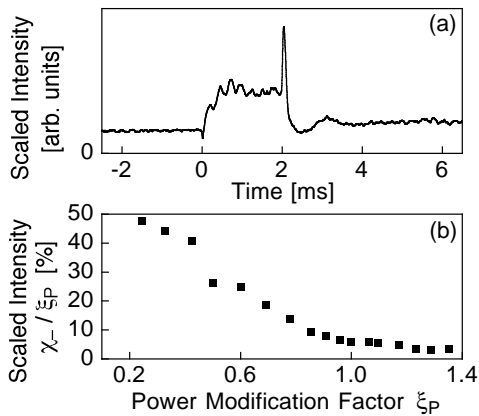}
\caption{ \label{Fig3}
(a) Intensity $I_{-}$ plotted versus time. Between $t=0~ms$ and $t=2~ms$ the
total power coupled to the cavity is reduced by $50 \%$.
(b) Plot of the scaled intensity $\chi_{-} / \xi_{P}$ versus the power
modification factor $\xi_{P}$.}
\end{figure}

We have probed the response of the system to a sudden change of
the total power $P_{0}$ coupled to the cavity, which is introduced
during a 2~ms long time window in a region well before the jump
occurs. The change of $P_{0}$ is parametrized in terms of the
remaining fraction $\xi_{P}$. In Fig.3(a) a typical example of the
response is shown for $\xi_{P} = 0.5$. As a first immediate
reaction to the reduction of $P_{0}$ the intra--cavity intensity
in the unlocked mode $I_{-}$ rapidly drops on a time scale given
by the cavity decay time ${\gamma_{c}}^{-1}$, as might be
expected. However, subsequently an increased steady state value of
$I_{-}$ is approached on a much slower time scale. At $t=2$~ms the
original value of $P_{0}$ is reestablished, i.e., $\xi_{P}$ jumps
back to $1$. After a fast transient increase, $I_{-}$ drops nearly
to its original value before the change of $P_{0}$ was introduced.
Note also the additional ripple at about 3 kHz, which amounts to
about six times the radial vibrational frequency. We have recorded
similar traces for varying values of $\xi_{P}$. In Fig.3(b) the
value of $\chi_{-} / \xi_{P}$ at $t=1.9$~ms is plotted versus
$\xi_{P}$, i.e. $I_{-}$ is scaled to $I_{0} \xi_{P}$. This graph
nicely illustrates the paradox behavior that more external power
yields less intra-cavity power.

Our system is classically described by $6N + 2$ coupled non-linear
differential equations for the $3N$ position and momentum coordinates and
the two complex electric field amplitudes of the cavity modes \cite{Gan:00}. We assume
a Gaussian geometry of the cavity modes with atoms located merely close to
the focus. We can combine the two equations for the fields into a single
equation because the servo lock keeps the field amplitude of the
locked mode at a constant value.
A significant part of the experimental findings can be explained by
a simple model based on the assumption of a thermal sample of atoms adiabatically
adjusting to the potential well depths and positions.
We may thus eliminate 6N equations of motion concerned with the external
degrees of freedom and remain with a single non-linear
differential equation for the intra-cavity complex electric field
amplitude $a(t)$ of the unlocked mode scaled to $\sqrt{I_{0}}$. Leaving the
details of its derivation to a forthcoming publication we merely indicate the result here:

\begin{eqnarray}
\frac{d}{d \tau} a \, = \,
i \frac{U N}{\sqrt{\chi_{0+}}} \, L(a) \, |a| a \,
- \, a  \, +   {}
\nonumber \\
+ \, \sqrt{\chi_{0-}} \, - \, i U N \, \sqrt{\chi_{0+}} \, L(a) \,
\frac{a}{|a|} \, \, ,
\nonumber \\
L(a) \equiv e^{-   \eta_{ax} \sqrt{\frac{|a_{0}|}{|a|}}   } \, \,
\frac{1}{1 + \eta_{rad} \, \frac{\sqrt{\chi_{0+}}\, + \, |a_{0}|}{\sqrt{\chi_{0+}} \, + \, |a|}}
\end{eqnarray}

In this equation $\tau \equiv \gamma_{c} t$ and $a_{0} \equiv a(t=0)$.
The function $L(a)$ is comprised of two factors describing
the axial and radial degrees of localization.
The truncation parameters $\eta_{ax}$ and $\eta_{rad}$ are defined as the
ratios between the thermal energy $k_{B} T$
and the well depth at time $t=0$ for the axial and radial
directions respectively. Eq. (1) can be readily integrated numerically, using the initial values
$a_{0}$ prepared by the MOT according to the traces of Fig.1(b).
In the time dependence of $N$ we account
for linear losses due to collisions with hot background atoms
and losses due to inelastic two-body collisions with rates measured in our
experiment. The values of $\eta_{ax}$ and $\eta_{rad}$ are determined
by temperature measurements to be approximately 0.5 and 0.3 respectively with about 0.1 uncertainty.
The difference is due to different well depths in axial and radial directions at t=0.

\begin{figure}
\includegraphics[scale=0.44]{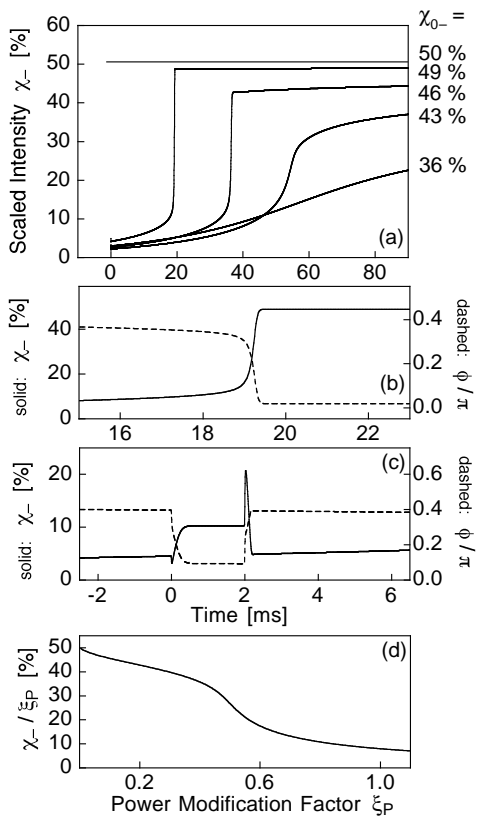}
\caption{ \label{Fig4} Theoretical simulations assuming adiabatic
atomic motion. The graphs in (a), (b), (c) and (d) correspond to
the experimental observations in Fig.1(b) Fig.1(c), Fig.3(a), and
Fig.3(b).}
\end{figure}

\begin{figure}
\includegraphics[scale=0.44]{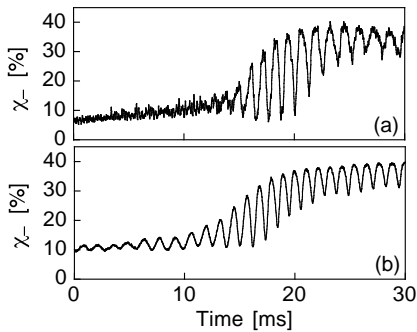}
\caption{ \label{Fig5}
(a) Experimental observation of squeezing oscillation.
(b) Simulation of the squeezing oscillations of (a) by means of solving the general equations of
motion for hundred atoms.}
\end{figure}

In Fig.4(a) and (b) we show simulations excellently matching our
experimental findings in Fig.1(b) and (c). For the initial
atom--cavity coupling strength $U N(t=0)$ we used the values 2.38,
2.23, 2.15, 1.75, which fall within a few percent of those
determined for the corresponding experimental traces, however
reduced by a common scaling factor 1.89. The need for this factor
is not surprising, because in our atom number measurements up to a
factor two uncertainty should be expected for the absolute values,
while relative values are on the few percent level. The basic
characteristics of the jump feature, i.e. the time for the jump to
occur and the rise time, are well reproduced. Eq. (1) lets us also
calculate the phase $\phi$ of the complex field $a(t)$ showing
that the jump is accompanied by a rapid shift of the optical
lattice of approximately a fraction $\lambda/10$, where $\lambda$
is the wavelength of the lattice light. This is illustrated by the
dashed trace in Fig.4(b). As shown in Fig.4(c) and (d) we may also
reproduce the findings of Fig.3. Again we calculated the
corresponding phase of $a(t)$, finding that also the change of
$P_{0}$ yields a spatial shift of the lattice.

We may not expect that our simple model can account for
non-adiabatic motion as for example the squeezing oscillations
shown in Fig.1(b) and Fig.5(a). Non-adiabatic motion is
particularly expected to occur with regard to the radial degrees
of freedom, which are not very tightly bound. In order to admit
non-adiabatic dynamics we have solved the complete system of
coupled equations of motion for hundred particles including axial
and radial degrees of freedom. The result shown in Fig.5(b) nicely
reproduces the frequency observed in (a). The calculation even
reproduces the fact that the frequency of the squeezing
oscillation slightly decreases with time, a detail that we cannot
explain so far. Apart from such squeezing oscillations the results
obtained by the full calculations are in good agreement with those
of the adiabatic model.

In summary, we have studied collective atomic motion and
corresponding non-linear phenomena arising in an optical lattice
formed inside a high finesse ring cavity. The spatial phase of the
lattice is not pinned by the phases of the incoupled laser beams,
but rather determined by the collective atom--cavity interaction.
The bistable behavior found experimentally for asymmetric pumping
can be understood in terms of an adiabatic approximation for the
atomic motion, whereas the general equations of motion must be
considered to model self--induced squeezing oscillations arising
in a certain parameter range. In this article we have only
discussed selected aspects of the system dynamics. A wealth of
further interesting phenomena could be studied, as for example
collective atomic recoil lasing (CARL) \cite{Bon:94} which is
expected to occur for unidirectional pumping of the cavity.

\begin{acknowledgments}
This work has been supported by Deutsche For\-schungs\-gemeinschaft (DFG) under
contract number $He2334/3-2$. We are grateful to Helmut Ritsch for discussions
and numerical support.
\end{acknowledgments}


\begin{thebibliography}{13}

\bibitem{Pin:00}
P.W.H. Pinkse, T. Fischer, P. Maunz, and G. Rempe, Nature {\bf 404}, 365-368  (2000).

\bibitem{Hoo:00}
C. Hood {\em et~al.}, Science {\bf 287}, 1457 (2000).

\bibitem{Doh:97}
A. C. Doherty, A. S. Parkins, S. M. Tan, and D. F. Walls, Phys.
Rev. A. {\bf 56}, 833 (1997).

\bibitem{Hor:97}
P. Horak {\em et~al.}, Phys. Rev. Lett. {\bf 79}, 4974 (1997).

\bibitem{Hec:98}
G. Hechenblaikner, M. Gangl, P. Horak, and H. Ritsch, Phys. Rev.
A. {\bf 58}, 3030 (1998).

\bibitem{Cha:03}
H.W. Chan, A.T. Black, and V. Vuletic, Phys. Rev. Lett. {\bf 90},
063003 (2003).

\bibitem{Els:03}
Th. Els\"{a}sser, B. Nagorny, and A. Hemmerich, Phys. Rev. A {\bf
67}, 051401(R) (2003).

\bibitem{Hem:99}
A. Hemmerich, Phys. Rev. A. {\bf 60}, 943 (1999).

\bibitem{Lam:95}
A. Lambrecht, E. Giacobino, and J.M. Courty, Opt. Commun. {\bf
115}, 199 (1995).

\bibitem{Nag:03}
B. Nagorny {\em et~al.}, Phys. Rev. A {\bf 67}, 031401(R) (2003).

\bibitem{Gan:00}
M. Gangl and H. Ritsch, Phys. Rev. A. {\bf 61}, 043405 (2000).

\bibitem{Bon:94}
R. Bonifacio, L. De Salvo, L. M. Narducci, and E. J. D'Angelo,
Phys. Rev. A {\bf 50}, 1716 (1994),  D. Kruse, C. von Cube, C.
Zimmermann, and Ph.W. Courteille, quant-ph/0305033 (2003).

\end{thebibliography}
\end{document}